\documentclass[conference]{IEEEtran}
\IEEEoverridecommandlockouts

\usepackage[utf8]{inputenc}
\usepackage[T1]{fontenc}
\usepackage{amsmath}
\usepackage{amssymb}
\usepackage{booktabs}
\usepackage{tabularx}
\usepackage{array}
\usepackage{cite}
\usepackage{url}
\usepackage{xcolor}
\usepackage{balance}
\usepackage{graphicx}
\usepackage[caption=false,font=footnotesize]{subfig}
\usepackage[hidelinks]{hyperref}

\newcommand{\cpp}{C\nobreakdash++}
\newcommand{\NA}{--}

\newcommand{\scigmark}{SciGMark}
\newcommand{\scimark}{SciMark}
\newcolumntype{Y}{>{\raggedright\arraybackslash}X}
\newcolumntype{C}{>{\centering\arraybackslash}p{0.82cm}}

\begin{document}

\title{Progress in Benchmarking Generics for Mathematical Computation}

\author{
\IEEEauthorblockN{Daniel Pang and Stephen M. Watt}
\IEEEauthorblockA{Cheriton School of Computer Science, University of Waterloo\\
Waterloo, Canada\\
\texttt{\{daniel.pang,smwatt\}@uwaterloo.ca}}
}

\maketitle

\begin{abstract}
The original SciGMark benchmark adapted the SciMark numerical suite to measure the cost of generic programming in scientific computing.  In the twenty years since, generics have become ordinary features of mainstream languages, but their implementation strategies have diversified.  This paper reports SciGMark 1.5, a benchmark study of specialized and generic implementations in modern languages.  The study has three aims.  First, it examines the consequences of the wide variety of generic-realization strategies used in current widely used languages.  Second, it extends the benchmark toward symbolic computation by adding finite-field linear algebra, finite-field FFT, and a na\"{\i}ve Gr\"obner{} basis computation.  Third, it asks how the original floating-point scientific kernels behave in the new language settings.  The measurements cover Rust, Java, Go, and TypeScript for the main suite, with additional C++ and Julia measurements for the Gr\"obner{} basis benchmark.  The study also records representative output artifact sizes, since code generation and runtime packaging are part of the cost model of generic programming.
The results confirm that the cost of generic programming is not inherent in abstraction itself, but depends strongly on when type information is resolved, how arithmetic values are represented, and whether the compiler or runtime can recover specific operations in the inner loops.  Ahead-of-time monomorphization usually makes generic code close to specialized code in the numerical tests.  Erased or object-based generic arithmetic can introduce substantial overhead, especially in arithmetic- and allocation-intensive code.  Runtime specialization occupies an intermediate position, offering flexibility and good steady-state performance when type inference and representation choices are favourable.
\end{abstract}

\section{Introduction}

Mathematical computing has always made heavy use of abstraction.  Many algorithms are naturally described over rings, fields, vector spaces, modules, polynomial domains, or structured matrices, rather than over a single machine type.  Generic programming is therefore attractive: it allows one implementation of an algorithm to be used over several concrete domains, and it gives a programming-language analogue of the mathematical generality already present in the algorithm.

This attraction comes with a practical concern.  The inner loops of scientific programs are often simple and frequent enough that small overheads become important.  A generic addition or multiplication that becomes a function call, an indirect dispatch, a heap allocation, or a boxed-object operation will usually be far more expensive than the corresponding primitive arithmetic instruction.  The relevant question is therefore not whether abstraction is useful, but when a language implementation can make the abstraction disappear at the point where performance matters.

The original \scigmark{} paper studied this question by adapting the \scimark{} numerical benchmark suite to generic implementations and comparing specialized and generic versions in C++, C\#, Java, and Aldor~\cite{DraganWatt2005}.  That work was motivated by the observation that scientific algorithms are well suited to generic programming, but that the cost of generic abstraction had to be understood before the approach could be used confidently in numerically intensive codes.  It also observed a distinction that remains important: symbolic computation had already accepted parametric polymorphism, whereas numerically intensive computation was more cautious.

Since then, the language landscape has changed.  Generics are now routine in mainstream languages, and languages differ not merely in surface syntax but in the strategy used to realize generic code.  Some systems generate specialized code ahead of time.  Some erase type parameters and recover behaviour through uniform object representations or interfaces.  Some specialize at runtime under a just-in-time compiler.  These choices influence runtime performance, code size, compilation behaviour, portability, and the programming experience.

This paper reports an updated and extended benchmark study, based on Pang's thesis measurements~\cite{Pang2026}, which we refer to as \scigmark{} 1.5.  It has three immediate objectives.  The first is to examine the practical consequences of the broad range of generic implementation strategies now present in widely used languages.  The second is to enlarge the benchmark beyond floating-point numerical kernels by adding symbolic and exact-arithmetic cases: linear algebra over finite fields, FFT over finite fields, and a na\"{\i}ve Gr\"obner{} basis computation.  The third is to see how the original style of floating-point scientific computation behaves in the newer languages.  The intent is not to rank languages by absolute speed.  Absolute speed depends on implementation choices, runtimes, libraries, allocation behaviour, and measurement conditions.  Rather, we compare the relative cost of moving from specialized to generic versions within each language and ask which generic-realization strategies preserve performance for the classes of kernels considered.

The contributions of the paper are as follows.
\begin{itemize}
\item We revisit the \scigmark{} methodology in the setting of modern, widely used languages whose generic mechanisms include ahead-of-time monomorphization, erasure, hybrid dictionary mechanisms, and runtime specialization.
\item We add symbolic and exact-arithmetic content to the suite through finite-field linear algebra, finite-field FFT, and a na\"{\i}ve Gr\"obner{} basis benchmark.
\item We measure how the floating-point scientific kernels from the \scigmark{} tradition behave in Rust, Java, Go, and TypeScript, and we include \cpp{} and Julia measurements for the Gr\"obner{} basis benchmark.
\item We identify the principal trade-offs observed in the experiments: specialization stage, representation of arithmetic values, allocation behaviour, conformance model, and deployment size.
\end{itemize}

The remainder of the paper is organized as follows.  Section~\ref{sec:revisiting} recalls the original \scigmark{} question and states the updated one.  Section~\ref{sec:benchmarks} describes the benchmark suite and its generic variants.  Section~\ref{sec:method} gives the language platforms and measurement method.  Section~\ref{sec:results} summarizes the results.  Section~\ref{sec:confounding} discusses confounding factors and interpretation.  Section~\ref{sec:conclusion} concludes.

\section{From SciGMark to SciGMark 1.5}
\label{sec:revisiting}

The earlier \scigmark{} study adapted \scimark{} to test the cost of generics in scientific computing.  The benchmark replaced certain occurrences of concrete numeric types by type parameters and domain interfaces.  For example, operations over \texttt{double} values were rephrased over a type representing a field or ring, and complex arithmetic was expressed as a generic construction over the coefficient type.  The original paper also included polynomial multiplication over a finite field to represent another style of mathematical computation~\cite{DraganWatt2005}.

That work separated two questions that are still easily confused.  The first is whether generic programs can express the algorithms naturally.  The second is whether implementations can make them fast enough.  The first question is largely settled: modern languages can express a wide range of parameterized numerical and algebraic algorithms.  The second question remains dependent on the language implementation.

It is useful to distinguish three broad realization strategies.
\begin{enumerate}
\item \emph{Ahead-of-time monomorphization} generates code for concrete type instantiations before execution.  This is the familiar template model in \cpp{} and is also central to Rust generics.
\item \emph{Erasure or uniform representation} compiles generic code to operate through a common representation, often with interface dispatch, boxed values, or runtime casts.  Java and TypeScript provide examples of this style, though their details are quite different.
\item \emph{Runtime specialization} specializes methods during execution based on observed argument types.  Julia is the principal example in this study.
\end{enumerate}
Go occupies a mixed position, using shape-based compilation and dictionary-like mechanisms in its implementation of generics.  This classification is intentionally coarse.  The point is not to reduce each language to one label, but to make visible the stage at which useful type information becomes available to the optimizer.

Table~\ref{tab:strategies} summarizes the dimensions that are most relevant to this study.  The table records where the cost of generic arithmetic is expected to appear.  A monomorphizing implementation pays by generating additional code and by depending on the compiler's ability to optimize each instantiated copy.  An erased implementation pays when primitive arithmetic must be represented by objects or by calls through an interface.  A runtime-specializing implementation pays when inference fails, when the program allocates values the optimizer cannot remove, or when compilation latency becomes part of the measured execution.

\begin{table*}[t]
\centering
\caption{Generic-realization strategies and principal cost mechanisms considered in the study.}
\label{tab:strategies}
\begin{tabularx}{\textwidth}{@{}lYYY@{}}
\toprule
Language & Realization strategy used here & Expected benefit & Expected source of cost \\
\midrule
Rust & Ahead-of-time monomorphization through traits and generics & Concrete operations can often be inlined and optimized & Extra abstraction layers, allocation choices, and missed optimizations in some domains \\
\cpp & Template-based ahead-of-time monomorphization & Mature native-code optimization and direct representation control & Template instantiation cost, code growth, and sensitivity to implementation details \\
Java & Type erasure with interface-based arithmetic objects and JIT optimization & Portable bytecode, mature runtime, small class artifacts & Boxing, allocation pressure, indirect dispatch, and uniform object representation \\
Go & Hybrid implementation using shape-based compilation and dictionary mechanisms & Simpler deployment model and moderate specialization & Interface indirection, dictionary passing, and representation choices \\
TypeScript & Compile-time erasure to JavaScript executed by Node.js & Structural typing and flexible source-level programming & Object wrappers, dynamic dispatch, allocation, and lack of efficient primitive generic numeric types \\
Julia & Runtime specialization through multiple dispatch and JIT compilation & Type-stable generic code can approach specialized execution & Compilation latency, inference sensitivity, allocation, and garbage collection \\
\bottomrule
\end{tabularx}
\end{table*}

The updated question is therefore: for representative floating-point, finite-field, and symbolic kernels, how much performance is lost when moving from specialized implementations to generic implementations, and how does this loss relate to the generic-realization strategy of the language?

Rust, Java, Go, and TypeScript provide the main cross-language comparison for the floating-point and finite-field kernels.  The Gr\"obner{} basis benchmark additionally includes \cpp{} and Julia, which supply important monomorphizing and runtime-specializing points of comparison for symbolic computation.

Table~\ref{tab:coverage} records the measurement scope.  The scope is sufficient for the three questions above: it compares several modern generic mechanisms on the original style of floating-point kernels, adds exact finite-field variants of linear algebra and FFT, and includes a symbolic polynomial benchmark.  A later \scigmark{} 2.0 should complete the language-by-kernel matrix, but the present measurements already support comparison of the main implementation strategies studied here.

\begin{table}[t]
\centering
\caption{Coverage of the present SciGMark 1.5 measurements.}
\label{tab:coverage}
\begin{tabular}{@{}lccccc@{}}
\toprule
Language & MC & SOR & LU & FFT & GB \\
\midrule
Rust       & yes & yes & yes & yes & yes \\
Java       & yes & yes & yes & yes & yes \\
Go         & yes & yes & yes & yes & yes \\
TypeScript & yes & yes & yes & yes & yes \\
\cpp       & --  & --  & --  & --  & yes \\
Julia      & --  & --  & --  & --  & yes \\
\bottomrule
\end{tabular}
\end{table}

\section{Benchmark suite}
\label{sec:benchmarks}

The benchmark suite keeps the spirit of \scigmark{} by comparing pairs of implementations.  A specialized implementation uses concrete arithmetic representations directly.  A generic implementation expresses the same algorithm through a parameterized domain abstraction.  We use the speedup ratio
\[
    \hbox{generic speedup} =
    \frac{\hbox{time of specialized implementation}}
         {\hbox{time of generic implementation}}.
\]
Thus a value near 1 indicates that the generic version is close to the specialized one.  Values below 1 indicate generic overhead.  Values above 1 occur occasionally because of measurement noise or secondary implementation effects.

The suite contains five kernels, summarized in Table~\ref{tab:benchmarks}.  Monte Carlo integration provides a simple floating-point baseline with little arithmetic abstraction in the hot loop.  Successive over-relaxation (SOR) stresses repeated array access and scalar arithmetic.  LU factorization provides a dense cubic-time kernel, measured over floating-point and finite-field domains.  FFT stresses complex arithmetic, memory movement, and finite-field variants.  The na\"{\i}ve Gr\"obner{} basis computation introduces exact arithmetic, polynomial operations, irregular control flow, and allocation.  Together these kernels separate three concerns: the behaviour of familiar floating-point scientific code in newer languages, the cost of exact finite-field arithmetic in generic form, and the additional representation pressure introduced by symbolic polynomial computation.

\begin{table}[t]
\centering
\caption{Kernels used in the benchmark suite.}
\label{tab:benchmarks}
\begin{tabularx}{\columnwidth}{@{}lYY@{}}
\toprule
Kernel & Domain & Main pressure in the generic version \\
\midrule
Monte Carlo & Floating point & Simple arithmetic and function-call overhead \\
SOR & Floating point & Array access, scalar arithmetic, memory layout \\
LU & Floating point and finite fields & Dense arithmetic in cubic-time loops \\
FFT & Complex and finite fields & Structured arithmetic, memory movement \\
na\"{\i}ve Gr\"obner{} basis & Finite fields and monomials & Exact arithmetic, allocation, irregular control flow \\
\bottomrule
\end{tabularx}
\end{table}

The floating-point kernels are derived from the SciMark/SciGMark tradition~\cite{SciMark2,DraganWatt2005}.  They remain in the suite because the original question has not disappeared: scientific codes in modern languages still rely on dense loops, arrays, complex arithmetic, and iterative kernels.  The finite-field and Gr\"obner{} basis cases broaden the study toward symbolic computation.  This is important because symbolic workloads do not always behave like dense floating-point kernels.  They often allocate many short-lived objects, use exact arithmetic, compare structured exponents, and follow input-dependent control flow.  These operations make the cost of generic abstraction less separable from representation and memory management.

The benchmark separates three ways in which genericity can become expensive.  The first is the representation of values.  A generic element of a finite field may remain an unboxed integer in a monomorphized implementation, but may become an object containing that integer in an erased implementation.  The second is the representation of operations.  An addition in a specialized loop may be a machine instruction, while a generic addition may require a method call or indirect dispatch.  The third is the representation of aggregates.  A matrix or polynomial may contain contiguous primitive values in the specialized version but references to boxed values in the generic version.  These three costs often occur together, so the benchmark reports end-to-end ratios rather than trying to assign a separate time to each mechanism.

The finite-field variants are included because they are the smallest exact-arithmetic step from scientific kernels toward symbolic computation.  In LU factorization they give linear algebra over a finite field.  In FFT they give a structured transform over a prime field.  In both cases, replacing floating-point arithmetic by arithmetic modulo a prime changes the cost balance.  Modular arithmetic introduces reductions and exact integer operations, while generic field elements introduce a domain interface.  The comparison is therefore not simply double arithmetic versus generic double arithmetic; it also asks how well a language can express and optimize a non-primitive arithmetic domain.

The Gr\"obner{} basis benchmark is intentionally na\"{\i}ve.  It is not meant to compete with specialized computer algebra systems.  Its purpose is to stress generic polynomial and monomial operations in a portable implementation.  The benchmark uses finite-field coefficients and compares two exponent-vector representations: a vector representation and a bitpacked representation. The vector representation is simpler and allocation-friendly in some languages.  The bitpacked representation is more compact and can improve locality, but it may require integer operations that are not equally natural across platforms.  This pair of representations is useful because it shows that the cost of genericity interacts with data layout.

The implementations are deliberately straightforward.  They do not call highly tuned numerical libraries or exploit special-purpose platform features.  This is a methodological choice.  The goal is not to measure the best possible FFT or LU factorization in each language.  The goal is to compare how similar algorithmic structures behave when the arithmetic domain is made generic.

The benchmark implementations are not identical at the level of machine representation.  They cannot be: the languages have different type systems, object models, numeric primitives, and runtimes.  The intended comparison is therefore within each language, specialized versus generic, and only secondarily across languages.

This choice also avoids a common failure mode in cross-language benchmarking.  If each language is allowed to use its best library, the result may measure library engineering rather than the language feature under study.  If each language is forced into a foreign idiom, the result may measure an artificial handicap.  The implementations used here occupy a middle position: they follow the same algorithmic structure and expose the same generic operations, while accepting the idiomatic representation mechanisms available in each language.

\section{Experimental method}
\label{sec:method}

Table~\ref{tab:platforms} gives the language implementations and compilation commands used in the study.  The experiments were run on a Nobara Linux 43 system with a 13th generation Intel Core i7-1370P processor, 14 visible cores, and 62 GiB of memory.

\begin{table}[t]
\centering
\caption{Language implementations and execution method.}
\label{tab:platforms}
\begin{tabularx}{\columnwidth}{@{}lY@{}}
\toprule
Language & Implementation and execution method \\
\midrule
Rust & Rust 1.92.0, compiled with \texttt{cargo build --release} \\
Java & Microsoft OpenJDK 25.0.1 LTS, \texttt{javac} 25.0.1 \\
Go & Go 1.25.5, compiled with \texttt{go build} \\
TypeScript & TypeScript 5.9.3, transpiled with \texttt{tsc}, executed by Node 24.12.0 \\
\cpp & \cpp{}17, GCC 11.4.0, compiled with \texttt{g++ -O3} \\
Julia & Julia 1.12.4, executed by the Julia runtime and LLVM JIT \\
\bottomrule
\end{tabularx}
\end{table}

Execution time was measured with the Unix \texttt{time} utility.  For kernels that otherwise complete as a single execution, multiple iterations were performed within one run to amortize setup costs and reduce the influence of just-in-time compilation.  LU factorization, FFT, and the na\"{\i}ve Gr\"obner{} basis benchmarks were executed ten times per run.  Monte Carlo and SOR use large iteration counts or problem sizes to obtain stable timings.

Input sizes were chosen to make the measurements large enough to expose steady behaviour while remaining executable on all relevant platforms.  This restriction matters because the most allocation-heavy generic implementations may run out of memory or require impractical time at sizes that are easy for a monomorphized implementation.  For the Gr\"obner{} basis benchmark, the inputs are cyclic-$n$ root systems for $n=4,5,6$.  The $n=6$ cases are large enough to expose representation and allocation effects without turning the benchmark into a test of only the fastest languages.

The principal performance measure is the specialized/generic speedup ratio within the same language, not absolute elapsed time across languages.  This reduces, but does not eliminate, the influence of language-specific baselines.  Absolute comparisons remain affected by compiler maturity, runtime engineering, memory layout, numerical representation, and the particular implementation decisions made in the benchmark.

We also measured a representative set of output artifacts.  Artifact size is not a direct measure of memory footprint, deployment cost, or compilation strategy in isolation.  A Java class file, a TypeScript-generated JavaScript file, a Julia source file, a dynamically linked \cpp{} executable, and a statically linked Go executable are not equivalent objects.  Nevertheless, artifact size is a useful secondary observation because monomorphization and runtime packaging both have visible effects on the generated output.

The measurements should therefore be read with two levels of comparison.  The first level is the within-language speedup ratio, which is the principal evidence about generic overhead.  The second level is the qualitative pattern across realization strategies.  A consistent slowdown in erased arithmetic across several kernels says something different from a single anomalous slowdown in one implementation. The analysis below gives more weight to repeated patterns than to isolated entries.

\section{Results}
\label{sec:results}

Table~\ref{tab:speedups} summarizes representative generic speedups for the largest commonly reported problem sizes.  The table should be read horizontally within a language and vertically only with care.  The principal pattern is that Rust, and where measured \cpp{}, usually remain close to specialized performance.  Java, Go, and TypeScript show larger overheads when generic arithmetic is forced through boxed objects, indirect operations, or less favourable runtime representations.  Julia is represented here only for the Gr\"obner{} basis benchmark and shows competitive, but not free, generic performance.

\begin{table*}[t]
\centering
\caption{Representative generic speedups, specialized time divided by generic time. Values near 1 indicate little generic overhead. Entries marked DNF did not finish within the practical limit used in the thesis measurements.}
\label{tab:speedups}
\begin{tabular}{@{}lcccccccc@{}}
\toprule
Language & MC & SOR & LU dbl. & LU $\mathbb{F}_p$ & FFT cplx. & FFT $\mathbb{F}_p$ & GB vec. & GB bits \\
\midrule
Rust       & 1.00 & 1.00 & 1.04 & 0.58 & 0.94  & 0.96  & 0.86 & 0.97 \\
Java       & 1.00 & 0.45 & 0.08 & 0.20 & 0.14  & 0.32  & 0.67 & 0.59 \\
Go         & 0.98 & 0.62 & 0.31 & 1.00 & 0.54  & 0.67  & 0.71 & 0.65 \\
TypeScript & 1.01 & 0.08 & DNF  & DNF  & 0.016 & 0.204 & 0.90 & 0.55 \\
\cpp       & \NA  & \NA  & \NA  & \NA  & \NA   & \NA   & 0.83 & 0.94 \\
Julia      & \NA  & \NA  & \NA  & \NA  & \NA   & \NA   & 0.84 & 0.93 \\
\bottomrule
\end{tabular}
\end{table*}

Monte Carlo integration behaves as a useful baseline.  At the largest tested input, all four fully implemented platforms show speedup ratios close to 1.  This indicates that not every use of a generic interface imposes a meaningful penalty.  When the hot loop can still be optimized to simple arithmetic and the abstraction is not repeatedly allocating or dispatching on complex values, the generic and specialized versions can be essentially indistinguishable.

SOR is more revealing.  Rust remains close to specialized performance, while Java, Go, and TypeScript show progressively larger generic overheads.  This is the first indication that the cost of abstraction is not uniform.  In a memory- and arithmetic-intensive loop, the representation of scalar values and the cost of accessing them repeatedly matter.  Java's erased generics and object representations lead to a substantial slowdown.  Go's overhead is moderate.  TypeScript's object-based numeric abstraction is particularly expensive in this benchmark.

LU factorization amplifies the pattern because the computation is cubic.  Rust remains close to specialized performance for floating-point LU, although the finite-field case has more visible overhead.  Java and TypeScript show large slowdowns in floating-point LU, and TypeScript did not complete the largest generic LU measurements.  Go is mixed: floating-point LU shows substantial overhead, but the finite-field LU measurements remain close to specialized performance.  This asymmetry is a useful warning against over-interpreting generic-realization strategy alone.  Details of domain representation and compiler treatment of the particular code can dominate the general expectation.

FFT gives a similar but clearer picture.  Rust remains within a small constant factor of its specialized versions for both complex and finite-field variants.  Java and Go incur larger overheads, and TypeScript's generic complex FFT is much slower than its specialized version.  The complex case is especially sensitive because a generic complex value introduces several temporary, potentally heap allocated,  values in a tight, repeated butterfly computation.  If the compiler cannot flatten and inline those operations effectively, generic abstraction becomes visible at every arithmetic step.

The na\"{\i}ve Gr\"obner{} basis benchmark changes the character of the workload.  It is symbolic, exact, allocation-heavy, and input-sensitive.  Rust, \cpp{}, and Julia show small to moderate overheads, especially for the bitpacked exponent representation.  Java and Go show larger overheads on the largest cyclic-6 inputs, and TypeScript is mixed: it is close to specialized in the vector-exponent case but significantly slower in the bitpacked case.  This mixed behaviour is unsurprising.  In symbolic workloads, the arithmetic abstraction is only one part of the cost.  Allocation, hashing or equality mechanisms, exponent representation, dynamic dispatch, and memory locality all contribute.

The language-level pattern is consistent with the strategy classification, but not reducible to it.  Rust is the cleanest example of the expected monomorphization result: generic code usually remains close to specialized code, and the largest slowdowns occur in cases where representation and allocation decisions matter in addition to dispatch.  \cpp{} shows similar behaviour in the Gr\"obner{} basis benchmark, though the present data do not yet cover the full numerical suite.  Julia's Gr\"obner{} measurements show that runtime specialization can be competitive on type-stable code, but also that dynamic specialization does not automatically remove all overhead in allocation-heavy symbolic computations.

Java exposes the cost of representing arithmetic domains through objects and interfaces.  Its generic code benefits from a mature JIT, but erased generics do not give primitive specialization of the field elements used in these benchmarks.  The result is a repeated penalty in kernels where generic arithmetic occurs inside tight loops.  Go occupies a middle position: it avoids some of the worst erased-object costs but still pays for its generic mechanism and representation choices in several kernels.  TypeScript is the least favourable platform for arithmetic-heavy generic abstractions in this suite, especially where generic numeric values become JavaScript objects rather than primitive numbers.

These observations should not be read as language verdicts.  They identify the cost model of the benchmark implementations.  A production library might change the result by using specialized storage, generated code, WebAssembly, native extensions, or a different representation of field elements.  That is precisely why the benchmark is useful: it shows where such engineering becomes necessary if one wants generic mathematical code to be fast.

Table~\ref{tab:artifacts} gives the measured artifact sizes for the Gr\"obner{} basis implementations and, where available, for the full test suite.  These numbers are useful mainly as a reminder that performance is not the only dimension of a generic-realization strategy.  Source-level or bytecode-level artifacts may be small because substantial functionality resides in an external runtime.  Statically linked binaries may be larger because they carry more of their execution environment.

\begin{table}[t]
\centering
\caption{Representative output artifact sizes.  Sizes are not directly comparable deployment footprints, but indicate the packaging consequences of the different systems.}
\label{tab:artifacts}
\begin{tabular}{@{}lrrrr@{}}
\toprule
Platform & Vec. & Bits & Gen. & Full suite \\
         & KB & KB & KB & KB \\
\midrule
Rust       & 527.58 & 496.96 & 572.73 & 836.73 \\
Java       &  20.02 &  17.87 &  49.75 & 159.37 \\
Go         & 2281.54 & 2297.46 & 2365.33 & 2607.96 \\
TypeScript &  18.51 &  10.75 &  39.71 & 121.86 \\
\cpp       &  65.25 &  57.24 & 115.87 & \NA \\
Julia      &  12.75 &  15.08 &  27.95 & \NA \\
\bottomrule
\end{tabular}
\end{table}

The artifact sizes broadly match expectations.  TypeScript and Julia distribute source-level artifacts in this measurement.  Java class files are small, but depend on the JVM.  \cpp{} produces compact dynamically linked native executables.  Rust produces larger native binaries, and Go produces much larger self-contained binaries because more of its runtime is embedded.  These observations do not contradict the performance results; they describe a different axis of the design space.

A useful way to read the results is to separate algebraic abstraction from representation abstraction.  Algebraic abstraction is the ability to write an algorithm over a field, ring, or polynomial domain.  Representation abstraction is the cost of representing the elements of that domain at runtime.  The first is desirable in all the systems studied here.  The second is where the performance differences arise.  A finite-field element that remains a register-sized integer has a different cost model from one stored as a heap object, even if both versions implement the same mathematical interface.

Table~\ref{tab:mechanisms} summarizes the main mechanisms exposed by the measurements.  It is included to make the discussion operational: each mechanism identifies a place where an implementation can improve, and each is tested by at least one kernel in the suite.

\begin{table*}[t]
\centering
\caption{Mechanisms by which generic abstraction becomes visible in the benchmark suite.}
\label{tab:mechanisms}
\begin{tabularx}{\textwidth}{@{}lYY@{}}
\toprule
Mechanism & How the suite exposes it & Consequence for interpretation \\
\midrule
Value representation & Double, finite-field, complex, and polynomial elements are used through generic domains & A slowdown may reflect boxing or loss of primitive representation rather than dispatch alone \\
Operation dispatch & Field operations occur repeatedly inside SOR, LU, FFT, and polynomial reduction loops & Small per-operation costs become large when multiplied through dense or iterative kernels \\
Aggregate layout & Matrices, arrays of complex values, polynomials, and exponent vectors differ in locality and indirection & A generic container may change cache behaviour even when the algorithmic loop is unchanged \\
Allocation pressure & Complex values, finite-field wrappers, monomials, and temporary polynomials may be short-lived & Garbage collection and allocator behaviour can dominate the apparent cost of genericity \\
Specialization timing & Monomorphized, erased, and JIT-specialized systems resolve type information at different stages & Similar source-level programs can present very different information to the optimizer \\
Artifact packaging & Native binaries, bytecode, source-level artifacts, and embedded runtimes package code differently & Runtime performance and output size should be treated as separate measurements \\
\bottomrule
\end{tabularx}
\end{table*}

This distinction also explains why Gr\"obner{} basis timings are not merely a symbolic analogue of the numerical timings.  In the numerical kernels, the generic operation is usually a scalar arithmetic operation embedded in a regular loop.  In the Gr\"obner{} basis benchmark, the generic operation is embedded in polynomial reduction, monomial comparison, exponent manipulation, and dynamic data-structure updates.  The cost of the field interface is still present, but it competes with allocation, equality tests, ordering, and memory locality.  The benchmark therefore tests whether a language can support generic mathematical objects as data structures, not only as scalar values.

There is a corresponding implication for library authors.  A language with efficient generic scalar arithmetic may still require specialized storage for matrices, polynomials, or exponent vectors.  Conversely, a language with relatively expensive generic dispatch may perform acceptably if the kernel is dominated by coarse-grained operations or if the runtime can specialize the relevant calls after warm-up.  The benchmark is meant to reveal these thresholds.  It is not enough to ask whether a language has generics; the question is where the generic boundary lies relative to the hot loop.

\section{Confounding factors and interpretation}
\label{sec:confounding}

The results should be interpreted as evidence about observed implementations, not as pure measurements of language semantics.  Generic realization strategy is important, but it is not the only variable.  Memory layout, allocation behaviour, cache locality, bounds checks, object representation, inlining decisions, garbage collection, and warm-up behaviour can all influence the measured times.  The benchmark design attempts to keep the algorithmic structure similar across languages, but no cross-language benchmark can make implementation details disappear.

This is why the specialized/generic ratio is the primary measure.  It asks a local question: given one language and one implementation style, how much is lost when the same algorithm is expressed generically?  This ratio is more meaningful than asking whether, for example, a Rust implementation is absolutely faster than a Java or Go implementation in a particular benchmark.  Even the ratio, however, can be affected by secondary differences.  A specialized implementation may use a representation that is not exactly mirrored by the generic one.  A generic implementation may require a wrapper type or interface object that changes memory layout.  A compiler may optimize one version more successfully because of small syntactic differences.

The just-in-time systems require additional care.  Java, TypeScript through Node.js, and Julia may change execution strategy during a run.  Looping several executions inside one timed run partially amortizes warm-up, but it does not provide a complete steady-state analysis.  A more detailed study would use profiling and runtime-specific counters to separate compilation time, optimization tier changes, allocation rate, garbage collection, and machine-code quality.

The benchmark set is also limited.  Five kernels cannot represent all scientific and symbolic workloads.  The suite intentionally emphasizes arithmetic-intensive code because that is where generic overhead is most visible.  Other applications may be dominated by I/O, communication, sparse data structure traversal, parallel scheduling, or library calls.  The conclusions should therefore be read as guidance for a class of mathematical kernels rather than as a universal statement about generics.

With these qualifications, the main interpretation is clear.  Abstraction is not itself the enemy of performance.  The cost appears when the language implementation cannot preserve or reconstruct the concrete representation needed in the hot loop.  Ahead-of-time monomorphization usually does this well.  Erasure and uniform object representations can do well when the generic layer is shallow, but can become costly when arithmetic values are boxed or operations are dispatched repeatedly.  Runtime specialization can be effective, but depends on type stability, inference, and the ability of the runtime to remove allocation and dispatch from the loop.

There is also a programming-language design lesson.  The performance model and the programming model are not independent.  Rust's explicit trait system, Java's explicit interfaces, Go's structural constraints, TypeScript's erased structural types, \cpp{} templates, and Julia's multiple dispatch all provide ways to write generic algorithms.  They differ in when errors are detected, how much conformance information is available to tools, and how directly the programmer can express algebraic requirements.  For mathematical software, this matters in addition to speed.  A generic library must be efficient, but it must also be maintainable and diagnosable.

The conformance model is especially important for algebraic code.  A field interface is not just a collection of function names.  It carries expectations about identities, inverses, exactness, and closure.  Most mainstream languages can express the required operations, but they vary in how much of the contract can be checked early.  Explicit traits and interfaces help tools report missing operations before execution.  Structural or implicit systems reduce annotation burden, but may defer errors until instantiation or execution.  Neither choice is intrinsically better; the point is that development cost is another dimension of the same design space as runtime cost.

The benchmark also indicates where more precise measurement is needed.  The present timings identify symptoms: a generic version is close to specialized, moderately slower, or much slower.  They do not by themselves explain whether the time went to dispatch, allocation, cache misses, garbage collection, missed vectorization, or JIT compilation.  A next measurement pass should therefore collect allocation counts, generated-code inspection where available, and hardware-counter data for the most informative cases.  This would turn the current high-level benchmark into a diagnostic benchmark.

\section{Conclusions and future work}
\label{sec:conclusion}

SciGMark was originally proposed to make the cost of generics in scientific computing visible.  Two decades later, the question remains relevant, but the answer is more differentiated.  Modern generic systems share enough expressive power to write a common family of numerical and symbolic kernels.  They do not, however, share the same performance consequences.  The important distinction is not generic versus non-generic code, but how the generic code is realized.

The results reported here show that generic abstractions can be close to free when the implementation specializes code while preserving concrete data representations.  They also show that erased, boxed, or highly dynamic representations can introduce large overheads in tight arithmetic loops.  Symbolic workloads complicate the picture: exact arithmetic, allocation, exponent representation, and irregular control flow can reduce the predictive power of any single generic-realization label.

The practical conclusion is that mathematical software should not treat generic programming as uniformly cheap or uniformly expensive.  It should be evaluated in the context of the intended language, workload, and deployment model.  For performance-critical numerical kernels, ahead-of-time specialization and predictable memory representation remain strong advantages.  For portable, reusable, or interactive systems, other trade-offs may be justified, provided their costs are understood.

There are several natural next steps.  First, the anomalous and mixed cases should be examined with profilers to separate allocation, dispatch, inlining and, especially, cache behaviour and garbage collection.  The most useful immediate targets are the TypeScript FFT and LU cases, the Java finite-field and Gr\"obner{} cases, and the Go cases where finite-field behaviour differs from floating-point behaviour.  These are not merely outliers; they are places where the benchmark can reveal which implementation mechanism is responsible for the measured ratio.

Second, the language-by-kernel matrix should be completed.  In particular, \cpp{} and Julia should be added for all tests, not only for the Gr\"obner{} basis benchmark, and Fortran and Aldor should be added across the full suite.  These additions are important for different reasons.  \cpp{} and Fortran anchor high-performance numerical computing.  Aldor connects directly to the original motivation from computer algebra and generic mathematical domains.  Julia represents a modern runtime-specializing scientific language whose performance depends strongly on type stability and generated code quality.  
We should also be explicit about how the different languages' inheritance mechanisms interact with the parameterization.
With these additions, the benchmark would be closer to the intended \scigmark{} 2.0 comparison.
We note that AI-assisted code generation could accelerate our investigation.  We have not yet done that for this project.

Third, the suite should be extended to parallel and heterogeneous settings.  Generic abstractions interact with vectorization, GPU kernels, tiling, and data movement in ways not captured by single-threaded scalar kernels.  A generic field element that is acceptable on a CPU may prevent SIMD vectorization.  A representation that is natural for a host language may be unsuitable for a GPU kernel.  These questions are now central to scientific computing and should be part of the next generation of the benchmark.

Finally, the benchmark itself should be packaged so that future language implementations can be tested as they evolve, in the same spirit as the original \scigmark{} suite.  The package should include fixed inputs, versioned implementations, scripts for collecting timings and artifact sizes, and a format for reporting specialized/generic ratios.  Without that discipline, each new comparison risks becoming a one-off experiment rather than a cumulative measurement of progress.

A useful reporting format should avoid a single aggregate score.  A single number would hide precisely the variation that the benchmark is intended to expose.  Each run should report the raw specialized and generic timings, the specialized/generic ratio, the problem size, compiler and runtime versions, optimization flags, warm-up policy, and hardware description.  For JIT-based systems, the report should distinguish first-run time from repeated-run or steady-state time where possible.  For artifact sizes, the report should state exactly what was measured: source files, bytecode, native executable, dynamically linked executable, or self-contained runtime image.

The next version should include a small set of diagnostic modes.  One mode should measure allocation counts or allocation volume when the runtime exposes them.  Another should permit inspection of generated code or intermediate representation in systems where this is practical.  A third should separate scalar field operations from aggregate data-structure operations.  A fourth should vary problem sizes systematically to identify when a kernel crosses memory hierarchy thresholds: private cache, shared cache, main memory bandwidth, and paging or garbage-collection pressure.  These transitions are likely to matter for matrices, FFT arrays, polynomial sets, and exponent-vector representations, and they may change the apparent cost of genericity.  These additions would keep the benchmark compact while making it more useful for implementers who want to improve generic performance rather than merely observe that it is slow.
\balance

\end{document}